\documentclass[aps,prl,twocolumn,amsmath,amssymb,showpacs,superscriptaddress,notitlepage,longbibliography]{revtex4-1}
\usepackage[colorlinks=true,linkcolor=blue,anchorcolor=red,citecolor=blue, urlcolor=blue]{hyperref}
\usepackage{bm}
\usepackage{graphicx}
\usepackage{color}

\begin{document}

\title{Forbidden Backscattering and Resistance Dip in the Quantum Limit as a Signature for Topological Insulators}

\author{Yiyuan Chen}
\affiliation{Shenzhen Institute for Quantum Science and Technology and Department of Physics, Southern University of Science and Technology, Shenzhen 518055, China}
\affiliation{Shenzhen Key Laboratory of Quantum Science and Engineering, Shenzhen 518055, China}

\author{Hai-Zhou Lu}
\email{Corresponding author. luhz@sustc.edu.cn}
\affiliation{Shenzhen Institute for Quantum Science and Technology and Department of Physics, Southern University of Science and Technology, Shenzhen 518055, China}
\affiliation{Shenzhen Key Laboratory of Quantum Science and Engineering, Shenzhen 518055, China}

\author{X. C. Xie}
\affiliation{International Center for Quantum Materials, School of Physics, Peking University, Beijing 100871, China}
\affiliation{Collaborative Innovation Center of Quantum Matter, Beijing 100871, China}

\date{\today}
\begin{abstract}
Identifying topological insulators and semimetals often focuses on their surface states, using spectroscopic methods such as angle-resolved photoemission spectroscopy or scanning tunneling microscopy. In contrast, studying the topological properties of topological insulators from their bulk-state transport is more accessible in most labs but seldom addressed. We show that, in the quantum limit of a topological insulator, the backscattering between the only two states on the Fermi surface of the lowest Landau band can be forbidden, at a critical magnetic field. The conductivity is determined solely by the backscattering between the two states, leading to a resistance dip that may serve as a signature for topological insulator phases. More importantly, this forbidden backscattering mechanism for the resistance dip is irrelevant to details of disorder scattering.
Our theory can be applied to revisit the experiments on Pb$_{1-x}$Sn$_x$Se, ZrTe$_5$, and Ag$_2$Te families, and will be particularly useful for controversial small-gap materials at the boundary between topological and normal insulators.
\end{abstract}

\maketitle


{\color{blue}\emph{Introduction}}. - Three-dimensional topological insulators \cite{Hasan10rmp,Qi11rmp,Shen17book} are characterized by topologically protected 2D surface states.
Most works focus on the surfaces, because they are also outposts for more exotic topological phases \cite{Yu10sci,Chang13sci,Fu08prl,Akhmerov09prl} that are of potential application in next-generation electronic devices.
However, in most samples, the bulk electrons overwhelm the surface electrons as the major carriers in the transport. Identifying topological insulators by detecting the bulk-state transport is an appealing topic of fundamental interest.

\begin{figure}[htbp]
\centering
\includegraphics[width=0.45\textwidth]{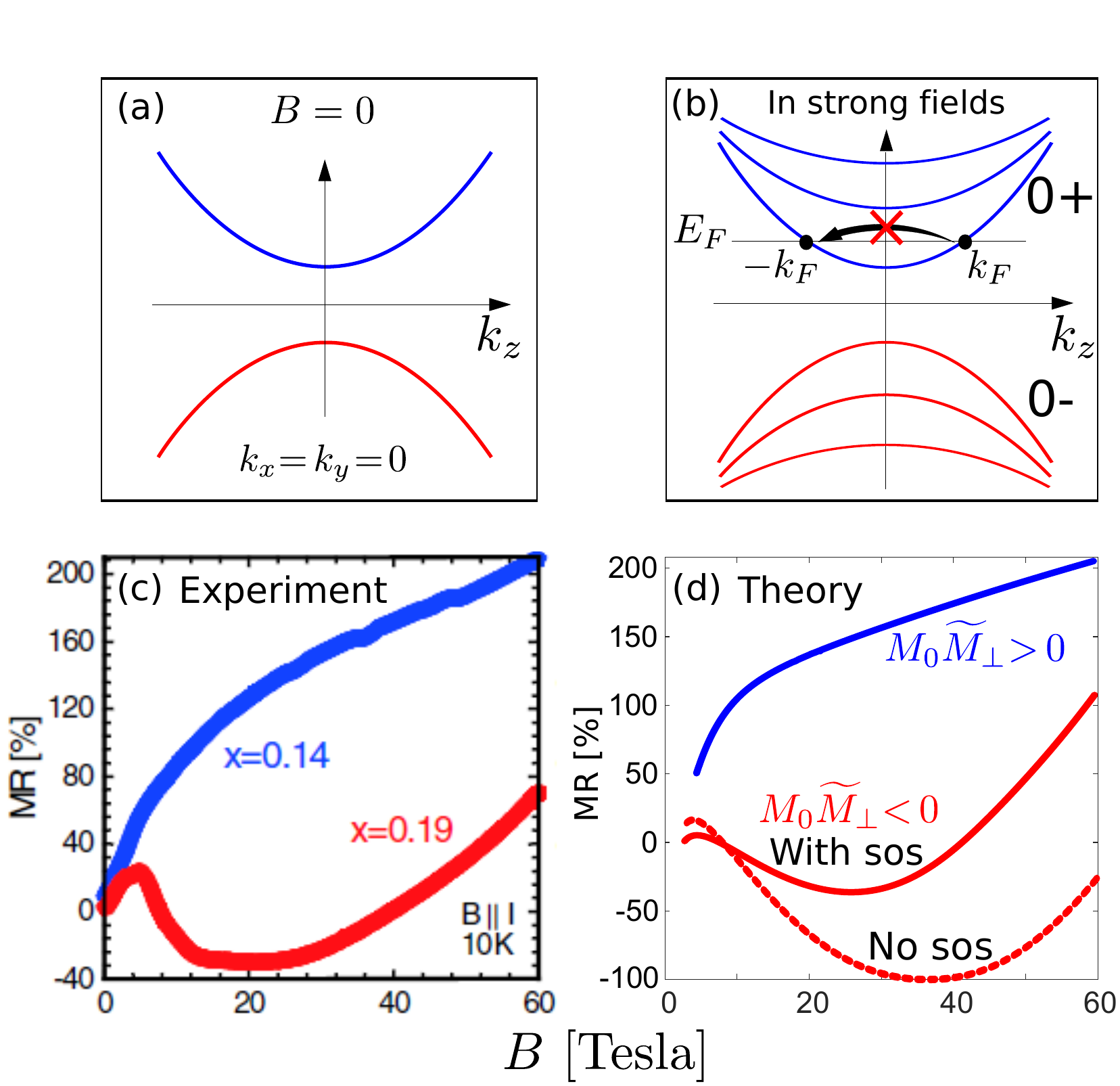}
\caption{In the quantum limit of a 3D topological insulator, the backscattering between the only two states at the Fermi energy can be forbidden at a critical magnetic field, leading to a resistance dip. (a) The zero-field energy spectrum vs $k_z$ of a 3D topological insulator at $k_x=k_y=0$. (b) In a strong magnetic field, the lowest Landau energy bands of the 3D topological insulator vs $k_z$. The Fermi energy $E_F$ crosses only the $0+$ Landau band. $k_F$ and $-k_F$ stand for the only two states at the Fermi energy. (c) The magnetoresistance of Pb$_{1-x}$Sn$_x$Se adapted from Ref. \cite{Assaf17prl}. (d) The calculated magnetoresistance using Eq. (\ref{Eq:rho-sos})  (see Sec. S2B of Ref. \cite{Supp} for details). The abbreviation ``sos" means spin-orbit scattering. The parameters are $M_0=-0.01$ eV, $M_z=0$, $\widetilde{M}_\bot$=18 eV\AA$^2$, $\alpha_1=100$ eVT, and  $\alpha_2=0.0025$ eVT$^{-2}$.\label{Fig:backscattering}}
\end{figure}

In a strong magnetic field, the 3D bulk states of a topological insulator quantize into 1D Landau bands. The lowest Landau band may inherit the topological information. In 2D, the crossing of the lowest Landau levels (not bands) has been used as a signature for the quantum spin Hall phase \cite{Konig07sci,Buttner11np}. But in 3D, how the lowest Landau band can be used to distinguish a topological insulator is seldom addressed. In this work, we study the resistance of a 3D topological insulator in parallel magnetic fields and when only the lowest Landau band is occupied, i.e., in the quantum limit [Fig. \ref{Fig:backscattering}(b)]. We find that for topological insulator phases, the backscattering in the quantum limit can be completely suppressed at a critical magnetic field. This suppression is irrelevant to the nature of the impurity scattering, and only depends on the spinor eigenstate of the lowest Landau band. As the backscattering is forbidden, the transport time diverges, and the resistivity shows a dip. This resistance dip resulting from the forbidden backscattering is absent in the trivial insulator phase, and thus can be used as a signature for the topological insulator phases. The mechanism of the forbidden backscattering is an eigenstate property and thus is new and different from the mechanism of Landau level crossing, which is a spectrum property. Also, this forbidden backscattering is absent in topological semimetals \cite{Lu15prb-QL,Goswami15prb,ZhangSB16njp}. Our theory is in good agreement with a recent experiment [Figs. \ref{Fig:backscattering}(c) and \ref{Fig:backscattering} (d)] and can be very useful for controversial small-gap materials at the boundary between topological insulators and normal insulators.

{\color{blue}\emph{Model and the quantum limit}}. - We start with a well-accepted $k\cdot p$ Hamiltonian for the bulk states in a topological insulator \cite{Zhang09np,Shen17book,Nechaev16prb}
\begin{eqnarray}
  H_0=C_{\bm{k}}+\begin{bmatrix}M_{\bm{k}}&0&iV_nk_z&-iV_\bot k_-\\0&M_{\bm{k}}&iV_\bot k_+&iV_nk_z\\-iV_nk_z&-iV_\bot k_-& -M_{\bm{k}}&0\\iV_\bot k_+&-iV_nk_z&0&-M_{\bm{k}}\end{bmatrix}
\end{eqnarray}
where $M_{\bm{k}}=M_0+M_\bot(k_x^2+k_y^2)+M_zk_z^2$, $C_{\bm{k}}=C_0+C_\bot(k_x^2+k_y^2)+C_zk_z^2$, and $k_x,k_y,k_z$ are the wave vectors, $M_i$,$V_i$ and $C_i$ are model parameters. With both $M_0M_\bot <0$ and $M_0M_z<0$, the model describes a 3D strong topological insulator with topologically protected surface states at all surfaces. Although a simple $k\cdot p$ model, it has been shown effective in the theories that work well for experiments, such as giving proper descriptions for the topological surface states \cite{Lu10prb,Shan10njp,ZhangY10np} and explaining the negative magnetoresistance in topological insulators \cite{Dai17prl,Wang12nr,He13apl,Wiedmann16prb,Wang15ns}.

In a strong magnetic field $B$ along the $z$ direction, the energy spectrum quantizes into a series of 1D Landau bands [Figs. \ref{Fig:backscattering}(a) and \ref{Fig:backscattering} (b)].
The energies of the lowest two Landau bands, denoted as $0+$ and $0-$, are $E_{0\pm}=C_0+C_zk_z^2+C_\bot/\ell_B^2\pm\sqrt{m^2+V_n^2k_z^2}$,
where the magnetic length $\ell_B\equiv \sqrt{\hbar/eB}$, $-e$ is the electron charge, and the mass term
\begin{eqnarray}\label{Eq:m}
m=M_0+M_zk_z^2+ M_\bot/\ell_B^2.
\end{eqnarray}
The gap between the lowest Landau bands can be determined by $m$ with $k_z=0$.

In the following, we will focus on an electron-doped quantum limit where the Fermi energy crosses only the $0+$ Landau band, whose eigenstate is found to be
\begin{eqnarray}\label{Eq:LB0}
|0,+,k_x,k_z\rangle &=&\begin{bmatrix}
0\\
-i \sin (\theta/2)\\
0\\
\cos(\theta/2)
\end{bmatrix}
|0,k_x,k_z\rangle,
\end{eqnarray}
where we have defined
\begin{eqnarray}
\cos\theta \equiv \frac{-m}{\sqrt{m^2 + (V_n k_z)^2}} ,
\end{eqnarray}
and $|0,k_x,k_z\rangle $ denotes the state of a usual zeroth Landau level timing a plane wave along the $z$ direction \cite{Lu15prb-QL}.


\begin{figure}
  \centering
\includegraphics[width=0.47\textwidth]{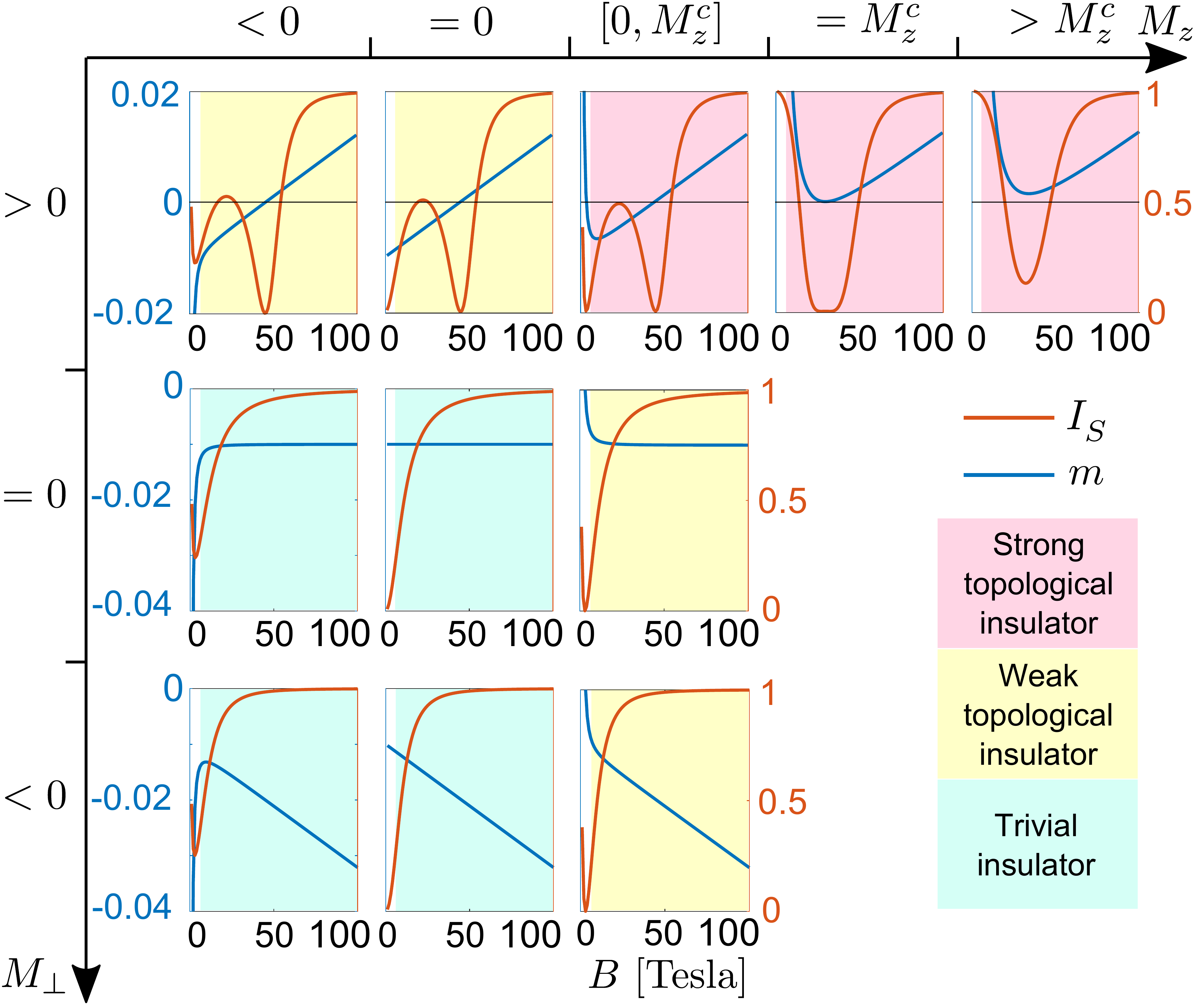}
  \caption{The mass term $m$ and form factor $I_S$ as functions of the magnetic field $B$ for $M_0<0$ and different $M_\bot$ and $M_z$. Red, yellow, and green backgrounds indicate the quantum limit for a carrier density of 6$\times$10$^{16}$/cm$^{3}$. Without loss of generality, we have assumed $M_0<0$, so $M_\bot>0$ and $M_z>0$ means strong a topological insulator, $M_\bot\le0$ and $M_z\le 0$ means a trivial insulator, $M_\bot\le0$ and $M_z>0$ means a weak topological insulator (001) and $M_\bot>0$ and $M_z\le0$ means a weak topological insulator (110). The parameters are $M_0$=-0.01 eV, $M_z$=-27, 0, 27, 820, 1200 eV\AA$^2$ from left to right, $M_\bot$=-13.5, 0, 13.5 eV\AA$^2$ from bottom to top.}\label{Fig:Mz}
\end{figure}

{\color{blue}\emph{Forbidden backscattering}}. -
The electronic transport in solids is more or less affected by the backscattering. In particular, the backscattering plays a dictating role in the present 1D Landau band, because there are only two states at the Fermi energy, as indicated by $k_F$ and $-k_F$ in Fig. \ref{Fig:backscattering}. The backscattering between these two states is characterized by the scattering matrix element between them. Using the spinor eigenstate in Eq. (\ref{Eq:LB0}), the modular square of the scattering matrix element between the $k_F$ and $-k_F$ states is found to be proportional to the form factor
\begin{eqnarray}\label{Eq:IS}
 I_S=\left.\cos^2\theta\right|_{k_z=k_F}.
\end{eqnarray}
$I_S$ vanishes when $m=0$, which means that the backscattering between state $k_F$ and state $-k_F$ is forbidden. According to Eq. (\ref{Eq:m}), $m$ vanishes at a critical magnetic field $B_c$ determined by
$M_0+M_zk_F^2+ M_\bot eB_c /\hbar=0$,
where $k_z$ becomes the Fermi wave vector $k_F$ at the Fermi energy.
For a topological insulator, $M_0M_z<0$ and $M_0M_\bot <0$, so $B_c$ has finite solutions at which the backscattering is completely suppressed. Later, we will show that this forbidden backscattering can lead to a dip in the resistance as a function of the magnetic field, which can be probed in experiments and can give a signature for topological insulator phases. We emphasize that, this forbidden backscattering is an eigenstate property and thus is new and different from the mechanism of Landau level crossing \cite{Konig07sci,Buttner11np}, which is a spectrum property.


{\color{blue}\emph{Zero point in form factor}}. - Now, we analyze the zero points of $I_S$, at which the backscattering is forbidden. For a fixed carrier density $n $, the Fermi wave vector $k_F$ depends on $B$ in terms of
$  k_F= 2\pi^2\hbar n /e B $ \cite{Lu15prb-QL}, so $m$ becomes
\begin{eqnarray}\label{Eq:m-B}
m=M_0+M_z(\frac{2\pi^2\hbar n }{e})^2\frac{1}{B^2}+ M_\bot \frac{e}{\hbar}B.
\end{eqnarray}
Depending on the signs of $M_0, M_z, M_\bot$, we have three phases:
(i) Strong topological insulator, $M_0M_z<0$ and $M_0M_\bot<0$. (ii) Weak topological insulator, $M_0M_z<0$ and $M_0M_\bot\ge 0$, or $M_0M_z\ge 0$ and $M_0M_\bot<0$. (iii) Trivial insulator, $M_0M_z\ge0$ and $M_0M_z\ge 0$.
Figure \ref{Fig:Mz} shows $m$ in Eq. (\ref{Eq:m-B}) and the corresponding $I_S$ as functions of $B$ for the three phases. Every time $m$ has a zero point, $I_S$ also has a zero point. For the trivial insulator phase, $I_S$ has no zero point. For both the weak and strong topological insulator phases, $I_S$ has one zero point in the quantum limit.
For a strong topological insulator, $I_S$ may have two zero points (Fig. \ref{Fig:Mz}, row 1, column 3), in which case our theory may not apply to the lower one, because it may not be within the quantum limit for realistic model parameters.
In addition, above a critical value $M_z^c$, there is no zero point (Fig. \ref{Fig:Mz}, row 1, columns 4 and 5). But the corresponding $M_z^c$ is too large (205 eV{\AA}$^2$ for the material we discuss) to be regarded as reasonable, e.g., $M_z=3.35$eV{\AA}$^2$, 9.25eV{\AA}$^2$, 22.12eV{\AA}$^2$ for Bi$_2$Se$_3$, Bi$_2$Te$_3$ and Sb$_2$Te$_3$ \cite{Nechaev16prb}. Considering $M_z$ is usually less that 100 eV\AA$^2$, it is safe to have the higher zero point of $I_S$.
Therefore, we conclude that the zero point of $I_S$ is only possible for either strong or weak topological insulator phases, in which the backscattering may be forbidden in the quantum limit.

\begin{figure}
  \centering
\includegraphics[width=0.47\textwidth]{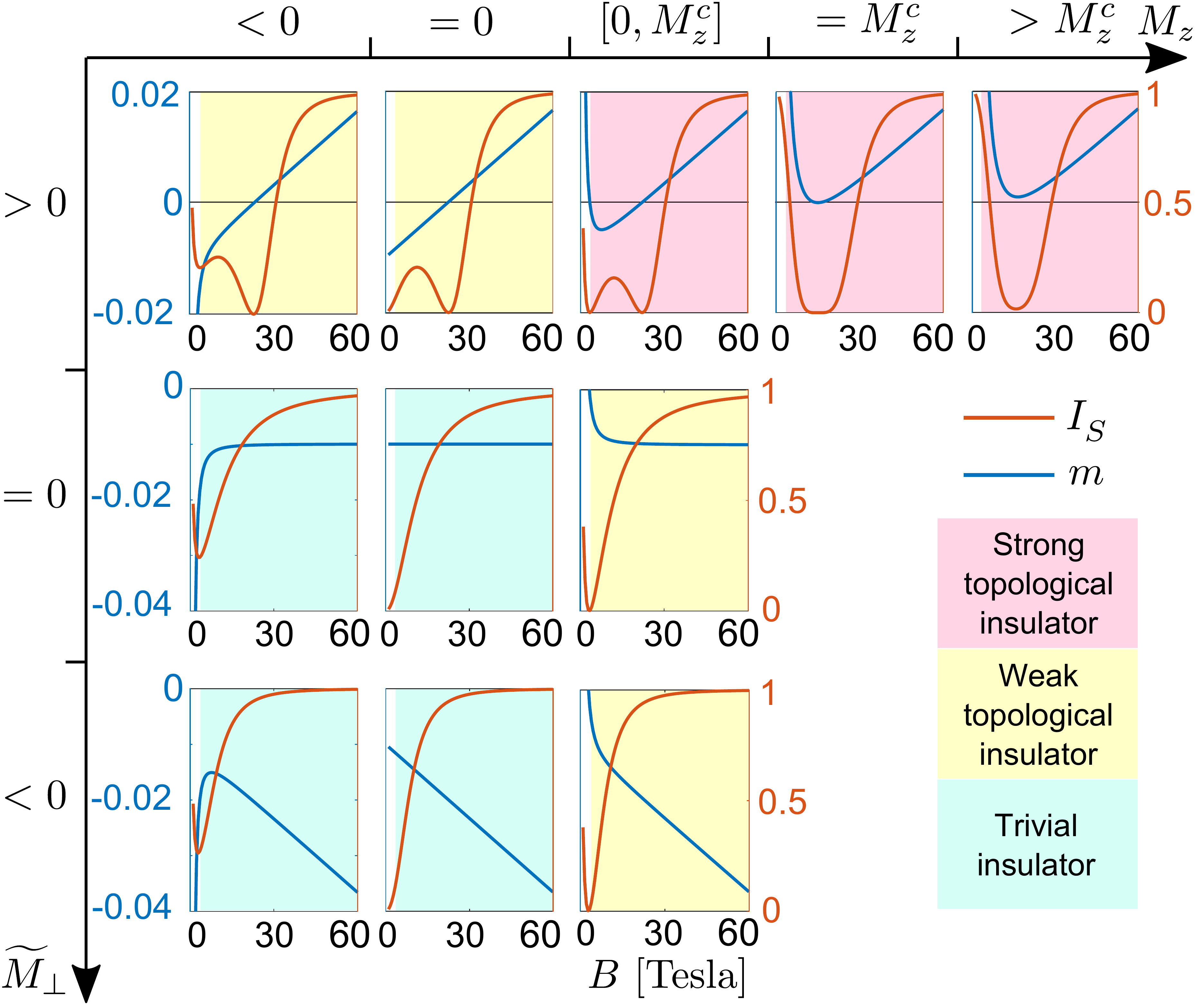}
  \caption{The same as Fig. \ref{Fig:Mz} but with the Zeeman effect, which shifts the zero point to a lower magnetic field. The parameters are the same as those in Fig. \ref{Fig:Mz}, except that $M_z$=-27, 0, 27, 205, 270 eV\AA$^2$ from left to right; $g_z^c$=7.5; and $g_z^v$=-7.5. }\label{Fig:Mz-Zeeman}
\end{figure}

{\color{blue}\emph{Zeeman effect}}. -
Above, we ignored the Zeeman effect, which we will show only quantitatively but not qualitatively changes the zero point of $m$ and $I_S$. The Hamiltonian of the Zeeman part reads
\begin{eqnarray}
  H_z=\frac{\mu_B}2\begin{bmatrix}g_z^vB_z&g_p^vB_-&0&0\\
  g_p^vB_+&-g_z^vB_z&0&0\\
  0&0&g_z^cB_z&g_p^cB_-\\
  0&0&g_p^cB_+&-g_z^cB_z
  \end{bmatrix},
\end{eqnarray}
where $\mu_B$ is the Bohr magneton and $g_{z,p}^{v,c}$ are $\mathrm{Land\acute{e}}$ $g$ factors for valance or conduction bands along the $z$ direction and in the $x$-$y$ plane, respectively. With the Zeeman effect, the mass term is corrected to
$m=M_0+M_zk_z^2+ M_\bot\frac{e}{\hbar}B+ B(g_z^c-g_z^v)\mu_B/4
$, so the Zeeman effect is to correct $M_\bot$ to  $\widetilde{M}_\bot=M_\bot  +(g_z^c-g_z^v)\mu_B \hbar/4e$. Figure \ref{Fig:Mz-Zeeman} shows that the Zeeman effect does not change the physical picture in Fig. \ref{Fig:Mz} much, but it shifts the zero point of $I_S$ to lower magnetic fields.

{\color{blue}\emph{Conductivity in the quantum limit}}. -
Now we show how the resistivity is dictated by the backscattering (see details in Ref. \cite{Supp}). Along the direction of the magnetic field, there is no Hall effect, so the resistivity is the inverse of the conductivity, i.e.,
$\rho_{zz}=1/\sigma_{zz}$.
In the quantum limit, the conductivity is contributed only by band $0+$, and can be explicitly expressed in terms of the transport time $\tau_{k_F}^\pm$ at the two states $\pm k_F$ on the Fermi surface \cite{Lu15prb-QL,ZhangSB16njp}
\begin{eqnarray}\label{Eq:sigma-zz-def}
\sigma_{zz} &=& \frac{e^{2}}{h}\frac{\hbar v_{F}\Lambda}{2\pi \ell_B^2}(\frac{\tau_{k_F}^+}{\hbar}+\frac{\tau_{k_F}^-}{\hbar}),
\end{eqnarray}
where $-e$ is the electron charge, $\Lambda$ is a value to correct the van Hove singularity at the band edge \cite{Lu15prb-QL}, and the Fermi velocity can be found as
$\hbar v_F=\left. \partial E_{0+}/\partial k_z\right|_{k_F}$.  The transport time $\tau_{k_F}^\pm$ can be found as \cite{ZhangSB16njp,Lu11prl}
\begin{eqnarray}\label{Eq:tau}
  \frac{\hbar}{\tau_{k_F}^\pm}&=&
  4\pi\sum_{k_x'k_z'}\langle|U_{0,0}^{k_z=\pm k_F,k_z'}|^2\rangle\frac{\Lambda\delta(k_z'\pm k_F)}{\hbar v_F},
\end{eqnarray}
where $k_z'=\mp k_F$ for $k_z=\pm k_F$, so the transport time depicts the backscattering from $k_F$ to $-k_F$, and vice versa.
$U_{0,0}^{\pm k_F,\mp k_F}$ are the scattering matrix elements between the states $k_F$ and $-k_F$, and $\langle ...\rangle $ means averaging over impurity configurations.
Assume a general form of random impurities $
U(\mathbf{r}) = \sum_i U(\mathbf{r}-\mathbf{R}_i)$,
where $\mathbf{R}_i$ are the positions of randomly distributed impurities, and the function $U(\mathbf{r}-\mathbf{R}_i)$ depicts the impurity potential energy. Using the eigenstate in Eq. (\ref{Eq:LB0}), the matrix element of the scattering between $k_F$ and $-k_F$ can be expressed as
\begin{eqnarray}\label{Eq:U-IS}
\langle|U_{0,0}^{k_z=k_F,k_z'=-k_F}|^2\rangle=I_S \langle|U_{0,0}|^2\rangle,
\end{eqnarray}
where $I_S$ is the form factor in Eq. (\ref{Eq:IS}) and the rest depends on the specific form of $U(\mathbf{r})$ in real space. Equation (\ref{Eq:U-IS}) shows that no matter the form of the real-space part, the scattering matrix element and the inverse of the transport time always vanish when $I_S$ vanishes. By understanding this and combining the above equations, the resitivitity can be expressed as
\begin{eqnarray}\label{Eq:sigma_zz-form}
  \rho_{zz}=I_S /\sigma_0,
\end{eqnarray}
where $\sigma_0$ is the conductivity independent of the spinor inner product part. For different types of scattering potential, $\sigma_0$ takes different forms. But the form factor $I_S$ in Eq. (\ref{Eq:sigma_zz-form}) dictates that a topological insulator always has a resistance dip, regardless of $\sigma_0$.

\begin{figure}
  \centering
  \includegraphics[width=0.46\textwidth]{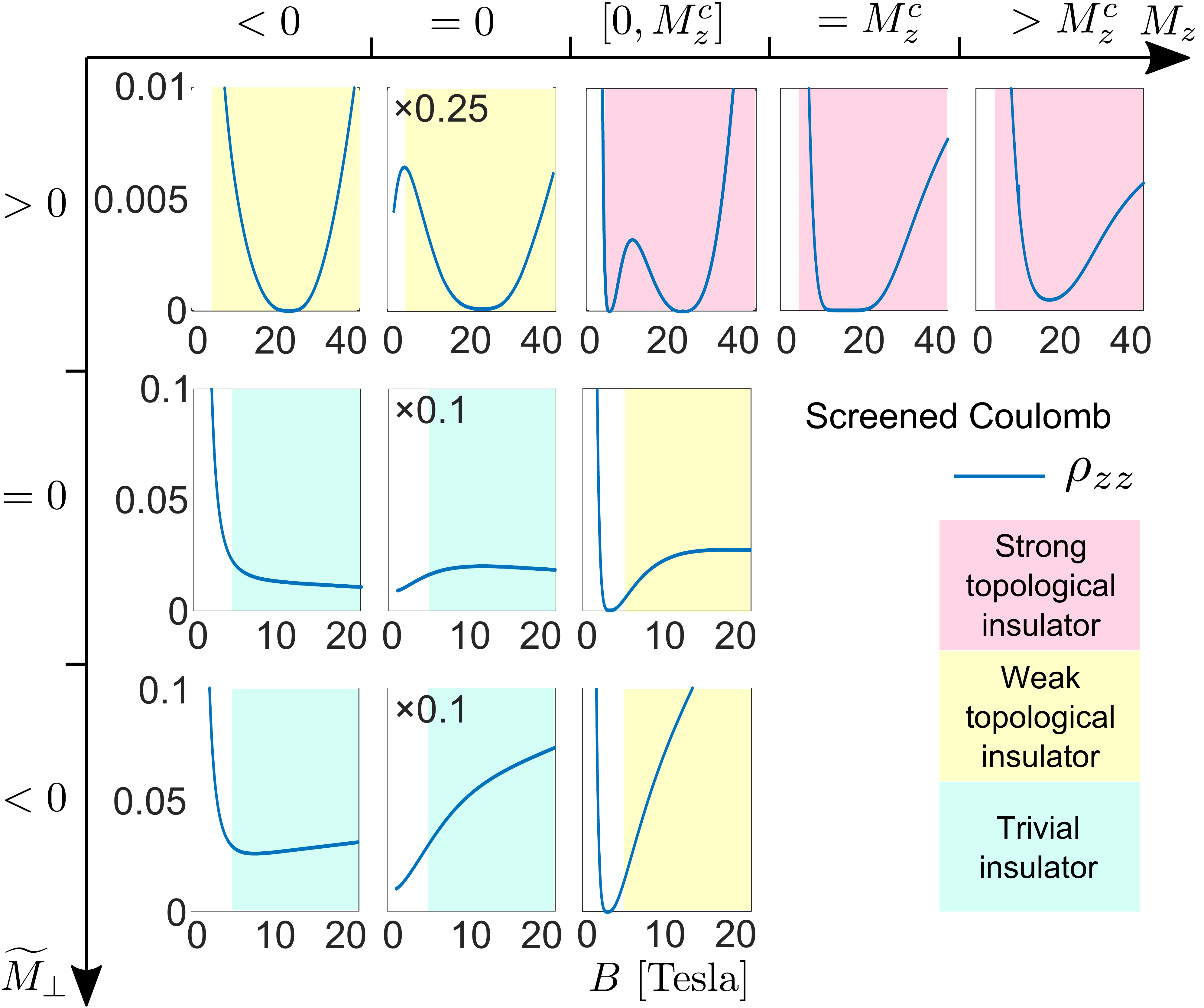}
  \caption{The same as Fig. \ref{Fig:Mz-Zeeman}, but for the resistivity $\rho_{zz}$ in the presence of the screened Coulomb scattering potential (in units of $2\pi^2 \hbar n_{\rm imp}/e^2\varepsilon$). $C_z$=1.409eV.\AA$^2$. $n_{\rm imp}$ for the trivial insulator and topological insulator are $6.8\times10^{18}$/cm$^3$ and $1.0\times 10^{19}$/cm$^3$, respectively. For $M_0>0$, the same results can be obtained by flipping $\widetilde{M}_\bot$ and $M_z$ accordingly. In the noncolored regions, the magnetic field is not strong enough to put all electrons into the lowest Landau bands.}\label{Fig:rho}
\end{figure}

We can test the argument with a systematic calculation of $\sigma_{zz}$, in the presence of the Gaussian \cite{ZhangSB16njp} and screened Coulomb \cite{Abrikosov98prb,Bruus04book} potentials, which are two common choices when describing the impurity scattering. For the Gaussian potential (see Ref. \cite{Supp} for details),
\begin{eqnarray}
\sigma_{0}=\frac{e^{2}}{h}\frac{(\hbar v_{F})^2(2d^2+\ell_B^2)}{  V_{\rm imp}\ell_B^2}   e^{4d^2 k_F^2} ,
\end{eqnarray}
where $d$ is the acting range of the Gaussian impurities, and $V_{\rm imp}$ measures the impurity density and scattering strength. As $d\rightarrow 0$, the conductivity reduces to
$  \sigma_0
= (e^{2}/h)
(\hbar v_F)^2 /V_{\text{imp}} $.
For the screened Coulomb potential (see Ref. \cite{Supp} for details),
\begin{eqnarray}\label{sigma-zz-sc}
\sigma_{0}
= \frac{e^{2}}{h}\frac{\hbar v_{F} \Lambda\varepsilon}{ \pi^2 n_{\text{imp}}e^2 \ell_{B}^4 },
\end{eqnarray}
where $\varepsilon$ is the dielectric constant and $n_{\rm imp}$ is the impurity density.
Figures \ref{Fig:rho} and Fig. S2 of Ref \cite{Supp} show the resistivity in the presence of the Gaussian and screened Coulomb potentials.
They both show clear dips in the resistivity for some weak topological phases ($M_z\le 0$ and $\widetilde{M}_\bot>0$ in row 1, columns 1 and 2) and strong topological insulator phases ($M_z\in[0,M_z^c]$ and $\widetilde{M}_\bot>0$ in row 1, column 3). More importantly, the positions of the minima on the $B$ axis does not change for different potentials.

{\color{blue}\emph{Spin-orbit scattering}}. - The spin-orbit scattering can improve the above picture and lead to a better fitting to the experiment. The spin configuration of our basis is $|\uparrow\rangle_1,|\downarrow\rangle_2,|\uparrow\rangle_3,|\downarrow\rangle_4$. We find that the off-diagonal spin-orbit scattering that couple 2 to 4 can change the result by lifting the exact forbidden backscattering (see Sec. S2B of Ref. \cite{Supp}). This lifting can lead to a better fitting to a recent experiment on the topological insulator material Pb$_{1-x}$Sn$_x$Se \cite{Assaf17prl}, which shows an unexpected suppression of the resistance in strong magnetic fields, as shown in Figs. \ref{Fig:backscattering} (c) and \ref{Fig:backscattering} (d). Specifically, in Fig. \ref{Fig:backscattering} (d), we include the spin-orbit coupling in the screened Coulomb scattering potential. As a result, the resistivity is found to be
\begin{eqnarray}\label{Eq:rho-sos}
  \rho_{zz}&=&\frac{h\pi }{e^3v_F^0\Lambda B}\left(\alpha_1\frac{I_S}{B}+\alpha_2(1-I_S)B^2\right),
\end{eqnarray}
where {\color{blue}the $\alpha_1$ term is due to the spin-independent scattering and diagonal spin-orbit scattering, and the $\alpha_2$ term is due to the off-diagonal spin-orbit scattering. In experiments, they can serve as fitting parameters.} By choosing proper $\alpha_1$ and $\alpha_2$, a $-40$\% change is obtained at the dip of the resistance, consistent with the experiment, as shown in Figs. \ref{Fig:backscattering} (c) and \ref{Fig:backscattering} (d).

{\color{blue}\emph{Experimental implications}} -
Although Pb$_{1-x}$Sn$_x$Se is claimed to be a topological insulator,
the interpretation in Ref.  \cite{Assaf17prl} adopted a theory for Weyl semimetal in delta scattering potentials \cite{Goswami15prb,Lu15prb-QL}, where the spinor wave function of the lowest Landau band takes the form
\begin{eqnarray}
|0,k_x,k_z\rangle &=&\begin{bmatrix}
0\\
|0\rangle
\end{bmatrix}
|k_x,k_z\rangle.
\end{eqnarray}
The spin part of the scattering matrix element is always 1, so a Weyl semimetal does not support the mechanism of the forbidden backscattering as the topological insulator does. In contrast, our theory gives a proper explanation and favors the experimentally observed resistance dip as a signature for weak topological insulators.

The resistance dip may not be observed in the Bi$_2$Se$_3$ family, because their large $M_0/\widetilde{M}_\bot$ ratios require inaccessible magnetic fields for the dip (Table S1 of Ref. \cite{Supp}).
Our mechanism of forbidden backscattering is irrelevant to the details of impurities,
and thus will be particularly useful for those controversial small-gap materials at the boundary between topological and normal insulators, such as the Ag$_2$Te \cite{Zhang11prl} and ZrTe$_5$ \cite{Weng14prx,Liu16nc} families, where controversies have built up over the years. In 2D, the quantum spin Hall phase may be probed by other approaches, such as interference effects \cite{Mani17srep}, but this has nothing to do with our proposal in 3D.

This work was supported by the Guangdong Innovative and Entrepreneurial Research Team Program (Grant No. 2016ZT06D348), the National Basic Research Program of China (Grant No. 2015CB921102), the National Key R \& D Program (Grant No. 2016YFA0301700), the
National Natural Science Foundation of China (Grants No. 11534001 and No. 11574127), and the Science, Technology and Innovation Commission of Shenzhen Municipality (Grant No. ZDSYS20170303165926217).


%

\end{document}